\documentclass[twocolumn,preprintnumbers,amsmath,amssymb,aps,prl,reprint]{revtex4}
\usepackage{graphicx}
\begin{document}

\title{
Enhanced Pinning For Vortices in Hyperuniform 
Substrates and Emergent Hyperuniform Vortex States  
} 
\author{
Q. Le Thien$^{1,2}$, D. McDermott$^{1,2}$, C.J. Olson Reichhardt$^{1}$, and  C. Reichhardt$^{1}$ 
} 
\affiliation{
$^1$ Theoretical Division and Center for Nonlinear Studies,
Los Alamos National Laboratory, Los Alamos, New Mexico 87545, USA\\ 
$^2$ Department of Physics, Wabash College, Crawfordsville, Indiana 47933, USA 
} 
\date{\today}
\begin{abstract}
Disordered hyperuniformity is a state of matter which has isotropic liquid like properties
while simultaneously having crystalline like properties
such as little variation in the density fluctuations over long
distances.
Such states arise for the packing of photoreceptor cells in chicken eyes,
jammed particle assemblies, and in nonequilibrium systems. An open question is
what possible applications could
utilize properties of hyperuniformity.
One of the major issues for applications of type-II superconductors is how to
achieve high critical currents by preventing the motion or depinning of vortices, so there is great interest
in understanding which pinning site geometries will lead to the optimal pinning of vortices.
Here, using large scale computational simulations, we show that vortices in a type-II superconductor
with a hyperuniform pinning arrangement exhibit enhanced pinning compared to
an equal number of pinning sites with a purely random arrangement, and that the enhancement is robust over a
wide range of parameters.
The stronger pinning arises in the hyperuniform arrays due to the suppression of pinning density fluctuations,
permitting higher pin occupancy and the reduction of weak links that lead to easy flow channeling.
We also show that in general, in amorphous vortex states in the presence of either random or hyperuniform pinning arrays,
the vortices themselves exhibit disordered hyperuniformity due to the repulsive nature
of the vortex-vortex interactions.
\end{abstract}

\maketitle

\vskip2pc
{\it Introduction---}

Disordered hyperuniformity describes amorphous systems
which have both liquid and crystalline properties \cite{1,2}.
The amorphous nature of these systems means that they are isotropic, as opposed to
crystalline systems which break spatial symmetries and exhibit Bragg peaks.
Hyperuniform systems also show strong suppression of density fluctuations
out to long length scales, a crystal-like property, where
the density per unit cell is fixed at a constant value.  This is in contrast
to a random assembly
or Poisson distribution of particles that can have large density variations
since it is possible
for points to accumulate
in certain regions or to have
extended regions devoid of points.
In 2003, Torquato and Stillinger proposed that
there are disordered many body
systems in which density fluctuations are suppressed out to very long wave lengths, and that
such systems can be characterized as exhibiting disordered hyperuniformity \cite{1}.
Since then, hyperuniformity has been studied in a growing number of systems
including the ordering of photoreceptor cells
in chicken eyes \cite{3}, jammed particle assemblies \cite{4,5,6}, block-copolymer systems \cite{7},
near nonequilibrium critical points \cite{8,9,10},
and even in certain quantum systems \cite{11}.
An open question is what possible applications there are for
systems that exhibit disordered hyperuniformity.
There have already been some proposals along these lines,
such as the use of hyperunformity
to create photonic materials with complete band gaps \cite{12}.

Here, we show that pinning sites geometrically arranged to exhibit hyperuniformity
have superior pinning
properties compared to an equivalent number of randomly arranged pinning sites for
magnetic vortices in a type-II superconductor over a wide
range of magnetic fields, substrate strengths, and applied drives.
One of the major issues for applications of type-II
superconductors is that the onset of vortex motion
limits the value of the current that can be carried by
the sample in the superconducting state, since the vortex motion produces
dissipation through a voltage response \cite{13,14,15}.
There have
been intense efforts directed at understanding ways
to increase the pinning of vortices by adding defects to superconducting samples in order
to
locally suppress the
superconducting order parameter,
creating low energy regions that trap vortices \cite{15,16}.
To emphasize the importance of pinning,
a general rule of thumb is that doubling the critical current reduces
the cost of using these materials by half \cite{15}.  
Since adding defects to the sample can degrade 
the material response, there is a limit to
the number of pinning sites that can be added.
Therefore it is important to determine the best way to spatially distribute  
a fixed number of pinning sites to create the strongest pinning for a wide range of fields.
One method is to arrange
the pinning sites in crystalline lattices
\cite{17,18,19,20,21,22,23}, diluted ordered lattices \cite{24,25},
quasiperiodic arrangements \cite{26,27},
conformal arrangements \cite{28,29,30}, or gradient arrays \cite{31,32,33}.
Typically in systems with crystalline arrangements of pinning sites, the 
depinning threshold is
strongly enhanced over that of random pinning arrangements
only for matching conditions under
which the number of vortices is an integer multiple of the number of pinning sites, whereas
under non-matching conditions, the periodic pinning arrays have lower depinning thresholds
than random arrays since the high symmetry of the array allows easy 1D vortex flow
channels to form along
symmetry directions of the crystal \cite{20,28}. In order to achieve
strong pinning for a wide range of parameters,
it would be ideal to place the pinning sites in a geometry that
has reduced pinning density fluctuations,
similar to crystalline arrays, while simultaneously remaining isotropic in order to eliminate      
easy flow symmetry channeling effects.
This suggests that pinning geometries with disordered hyperuniformity
could have ideal properties for vortex pinning.

Another question is whether amorphous assemblies of vortices in the presence of
random pinning arrange themselves
in a hyperuniform state or a random state. Generally, vortex structures
in the presence of pinning are described as either being ordered, as in a Bragg glass state
where there are no dislocations in the vortex lattice \cite{N1}, or as
amorphous where numerous topological defects
are present \cite{N1,N2,N3}. Due to the repulsive interaction between vortices,
strong density
fluctuations are highly energetically costly,
which suggests that the amorphous vortex
structure may be hyperuniform in nature when vortex-vortex interactions are relevant, and
more random in nature when pinning or thermal effects dominate.
Since hyperuniform states are expected to occur for certain charged systems \cite{2}, pinned
amorphous vortex systems may be ideal places to seek emergent hyperuniformity.
We show that disordered hyperuniform vortex
states arise for vortices interacting with either hyperuniform or random pinning arrays, which
suggests that hyperuniformity is a general feature of pinned vortex systems.
There are many
techniques that have been used to visualize large amorphous vortex assemblies
\cite{34,35,36,37,L2,38,39,40,L1},
and it would be interesting to re-examine this data
to see if hyperuniform or random configurations occur.
Additionally, there are
a wide class of systems that have many similarities to amorphous vortices in the presence of pinning
which may also exhibit hyperuniformity, including charge-stabilized
colloids \cite{41}, Wigner crystals \cite{42}, and skyrmions in chiral magnets
\cite{43,44}.

\begin{figure}
\includegraphics[width=\columnwidth]{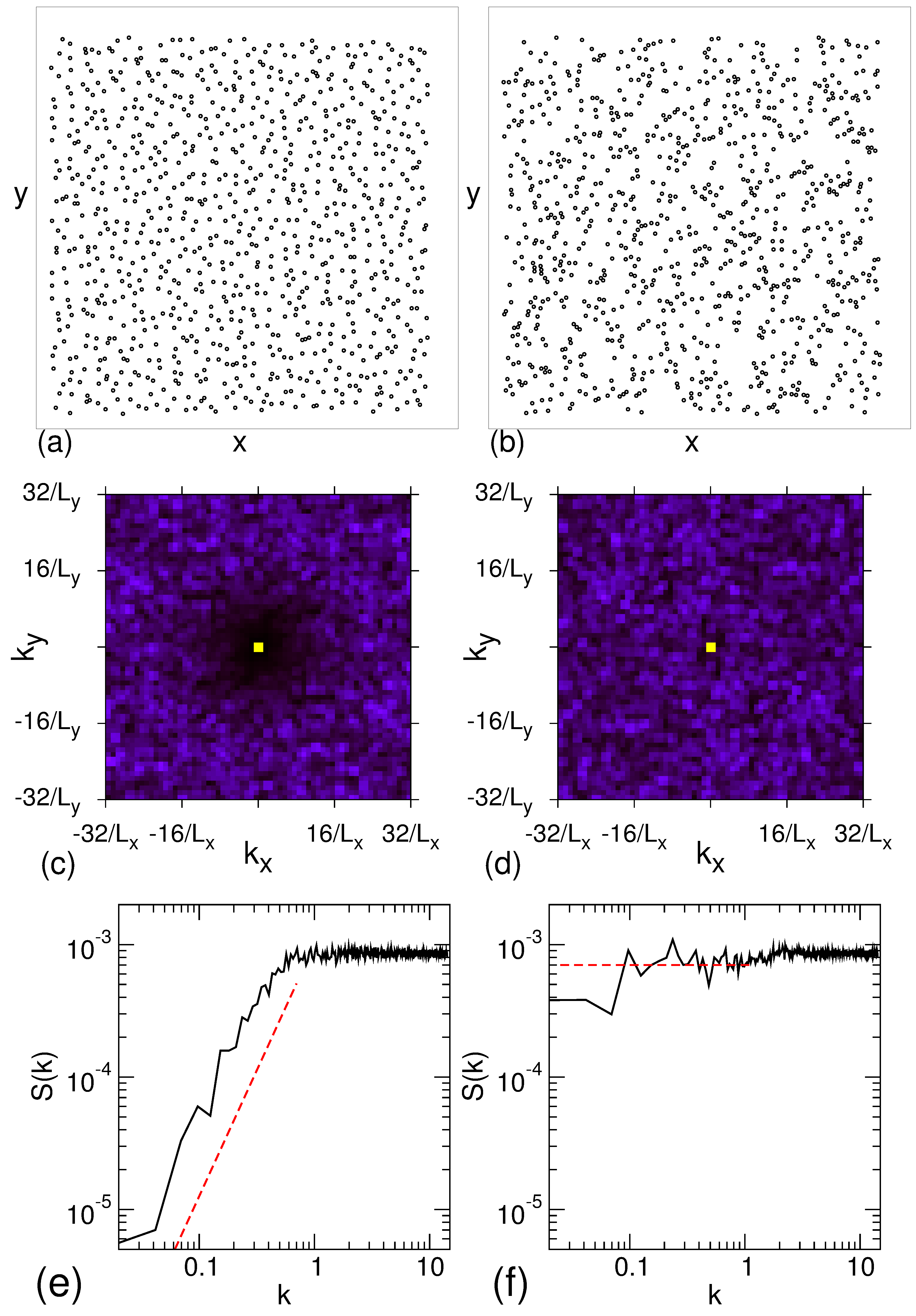}
\caption{The pinning site locations (open circles) for
  (a) a hyperuniform distribution and (b)
  a random distribution.
  (c) The corresponding structure factor $S({\bf k})$ for the hyperuniform
  pinning array showing that the weight vanishes at small ${\bf k}$
  and that the system is isotropic.
  (d) $S({\bf k})$ for the random pinning array,
  where the system is isotropic but the weight goes to a finite value  
  at small ${\bf k}$.
  (e) $S(k)$ vs $k$ for the hyperuniform array, where the dashed line is a fit
  to $S(k) \propto k^{1.9}$.
  (f) $S(k)$ vs $k$ for the random array
  showing that $S(k)$ goes to a constant value at small $k$.} 
\label{fig:1}
\end{figure}

{\it System description---}
The key feature of disordered hyperuniformity is the suppression of density
fluctuations out to long distances. This can be characterized in reciprocal space
by the behavior of the structure factor
$S({\bf k})=N_v^{-1}|\sum_i^{N_v} \exp(-i{\bf k} \cdot {\bf R}_i)|^2$, which goes to zero as $|{\bf k}|$ goes to zero, similar to a crystal; however,
unlike a crystal, the hyperuniform $S({\bf k})$ is isotropic and has no Bragg peaks
\cite{1,2}. In general, for a disordered hyperuniform system $S({\bf k})$ goes to zero
as $|{\bf k}|^{\alpha}$, where $\alpha$ is a
positive number.  For a random system, $S({\bf k})$ is
isotropic but it approaches a finite value
as $|{\bf k}|$ goes to zero.

We consider two-dimensional systems
in which we place $N_{p}$ pinning sites
arranged in either a random or a hyperuniform configuration
as shown in Fig.~\ref{fig:1}(a,b).  The pinning sites are modeled
as nonoverlapping local attractive parabolic wells
with radius $r_{p}$. The hyperuniform array is constructed by
setting up a square lattice of cells and placing one pinning site at a randomly chosen location within each cell \cite{1,2,CN}.
The random array is constructed using a Poisson distribution.
Figure~\ref{fig:1}(c,d) shows the corresponding
$S({\bf k})$ plots for the pinning configurations in Fig.~\ref{fig:1}(a,b).
For the random array,
$S({\bf k})$ has
constant weight at small ${\bf k}$,
while for the hyperuniform array, the weight vanishes at small ${\bf k}$.
In both cases $S({\bf k})$
is isotropic, indicative of an amorphous system.
In Fig.~\ref{fig:1}(e,f) we plot $S(k)$ versus $k$
for the hyperuniform and random arrays, showing that
$S(k)$ goes to zero at small $k$ for the
hyperuniform system, with the dashed line indicating a fit to $k^{1.9}$, while
for the random array $S(k)$ goes to a constant value at small $k$.
Within the sample we place $N_{v}$ vortices modeled as point particles with
a repulsion given by a pairwise Bessel function
$K_{1}(r)$ interaction as used in previous vortex simulations \cite{20,24,28,32}.
The initial vortex positions
are obtained by
starting from a high temperature state and cooling to $T= 0$
to obtain a ground state configuration.
After the initialization we apply a driving force, which experimentally corresponds to the
application of an external current that
creates a Lorentz force on the vortices.
We wait a fixed time at each drive increment to ensure that the
system has reached a steady state,
and then we measure the average vortex velocity
$\langle V\rangle=N_v^{-1}\sum_{i=1}^{N_v}{\bf v}_i \cdot {\bf \hat x}$
in the direction of the driving force
to determine when the vortices depin and to construct
velocity-force curves that are
proportional to experimentally measurable current-voltage curves.

{\it Numerical Methods---}
We utilize a particle model based on the London equations to
describe the superconducting vortices. 
The dynamics of a single vortex $i$
is governed by the following overdamped equation of motion:
\begin{equation}
\eta \frac{ d{\bf R}_{i}}{dt} =
{\bf F}^{vv}_{i} + {\bf F}^{vp}_{i} + {\bf F}^{D} ,
\end{equation}
where
${\bf v}_{i} = {d {\bf R}_{i}}/{dt}$ is the vortex velocity,
${\bf R}_{i}$ is the vortex position,
and $\eta$ is the damping term which is set to unity.
The interaction with the other vortices is repulsive and comes from the term 
${\bf F}^{ vv}_{i} = \sum^{N_v}_{j=1}
F_{0}K_{1}(R_{ij}/\lambda) \hat{\bf r}_{ij}$
where $F_{0} = \phi^{2}_{0}/2\pi\mu_{0}\lambda^3$,
$\phi_0$ is the elementary flux quantum, $\mu_0$ is the permittivity,
$R_{ij}=|{\bf r}_i - {\bf r}_j|$,
$\hat{\bf r}_{ij}=({\bf r}_i - {\bf r}_j)/R_{ij}$,
$K_{1}$ is the modified Bessel function
which falls off exponentially for large $R_{ij}$, and
$\lambda$ is the London penetration
depth which we set equal to $1.0$.
We place a cutoff on the interactions 
for vortex separations $R_{ij}/\lambda > 6.0$ for computational efficiency.
At $T = 0$ and in the absence of pinning,
the vortices form a triangular solid due to their mutually repulsive interactions.
The pinning force ${\bf F}^{vp}_{i}$
is produced by $N_p$ non-overlapping
harmonic
potential traps with a radius $R_{p}=0.15$
which can exert a maximum pinning force of $F_{p}$ on a vortex.
The driving term
${\bf F}^{D}=F_D{\bf \hat x}$
represents a Lorentz force from an externally applied current
interacting with the
magnetic flux carried by the vortices \cite{26}.
Our system is of size $L \times L$ with $L=36$,
and has periodic boundary conditions in the $x$ and $y$ directions.
The vortex density is $n_v=N_v/L^2$ and the pinning density is $n_p=N_p/L^2$.
In this work all forces are measured in units of $F_{0}$ and length in units of $\lambda$.

\begin{figure}
\includegraphics[width=\columnwidth]{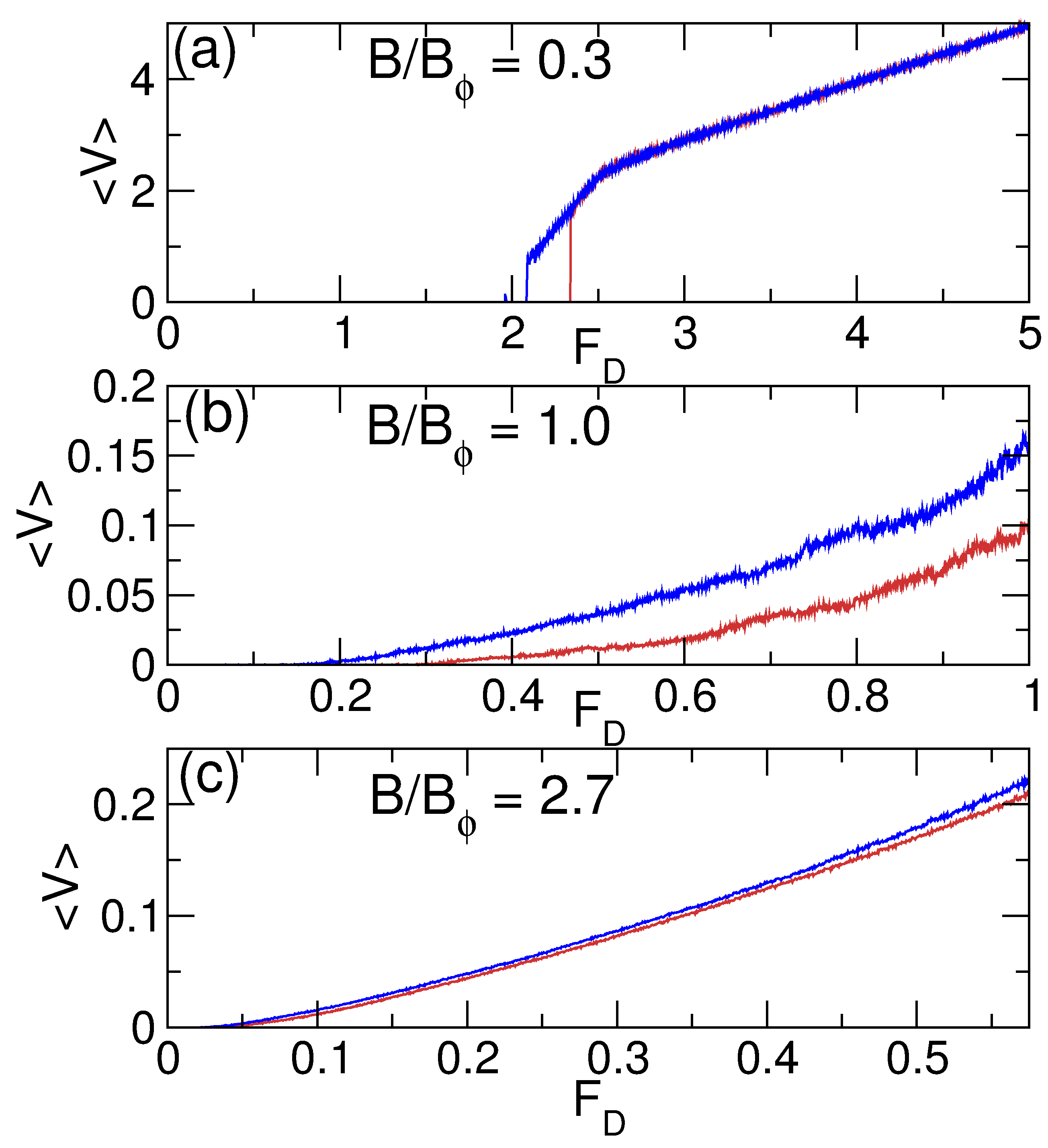}
\caption{ Vortex velocity $\langle V\rangle$
  vs driving force $F_{D}$ for a hyperuniform pinning array (red, lower curves) and
  random pinning array (blue, upper curves)
  for systems with pinning density $n_{p} = 0.7$ and
  pinning strength $F_{p} = 2.55$.
  (a) $B/B_{\phi} = 0.3$, where $B_{\phi}$ is the field at
  which there is one vortex per pinning site.
  The ratio of the depinning threshold for the hyperuniform array to
  that of the random array is $R = 1.125$.
  (b) At $B/B_{\phi}  = 1.0$, $R = 1.8$.
  (c) At $B/B_{\phi} = 2.7$, $R = 1.2$.
}
\label{fig:2}
\end{figure}

{\it Enhanced Pinning with Hyperuniform Substrates---}

In Fig.~\ref{fig:2} we plot the the vortex 
velocity $\langle V\rangle$ vs applied driving force $F_{D}$ for a system with
$N_p=900$ pinning sites at
a pinning density of $n_{p} = 0.7$
with $F_p=2.55$ arranged in either a hyperuniform or a random array.
The depinning threshold is defined to be the lowest value of $F_D$ for which
a persistent flow of vortices occurs
so that $\langle V\rangle > 0$.
Since the vortex density is proportional to the magnetic field,
we can define a matching field $B_{\phi}$ to be the field at which
there is exactly one vortex per pinning site.
At $B/B_{\phi} = 0.3$ in Fig.~\ref{fig:2}(a),
the depinning threshold for the hyperuniform array is $F^{\rm hyper}_{c}/F_{p} = 0.936$, while
that of the random array is $F^{\rm random}_c/F_p=0.832$.
We quantify the pinning enhancement $R$ as the ratio of these two
depinning thresholds, $R = F^{\rm hyper}_{c}/F^{\rm random}_{c}$,
obtaining $R=1.125$ in this case.
For $F_{D}/F^{\rm hyper}_{c} > 1.0$,
the velocity response for both pinning arrays is almost the same.  
In general, at lower fields
where the vortices are widely spaced, the vortex-vortex interactions are less relevant 
and the depinning threshold is dominated by the strength of the individual
pinning sites, so
in the extremely low field limit of a single vortex, $R  = 1.0$.
At $B/B_{\phi} = 1.0$, Fig.~\ref{fig:2}(b)
shows that the depinning threshold is more strongly enhanced by
the hyperuniform array and
$R = 1.8$.
Here, once both systems have depinned,
the velocity response for the random array is higher than
that of the hyperuniform array, indicating that even within
the sliding state, the hyperuniform array is more effective in reducing the dissipation.
For $B/B_{\phi} = 2.7$ in Fig.~\ref{fig:2}(c), there
is a reduced enhancement of
$R = 1.2$, and above depinning, the velocity response
of the hyperuniform array is slightly lower than that of the random array.
In general, at the higher vortex densities, the vortex-vortex interactions
begin to dominate over the vortex-pin interactions,
so the difference in the pinning effectiveness of the two pinning geometries is reduced.

\begin{figure}
\includegraphics[width=\columnwidth]{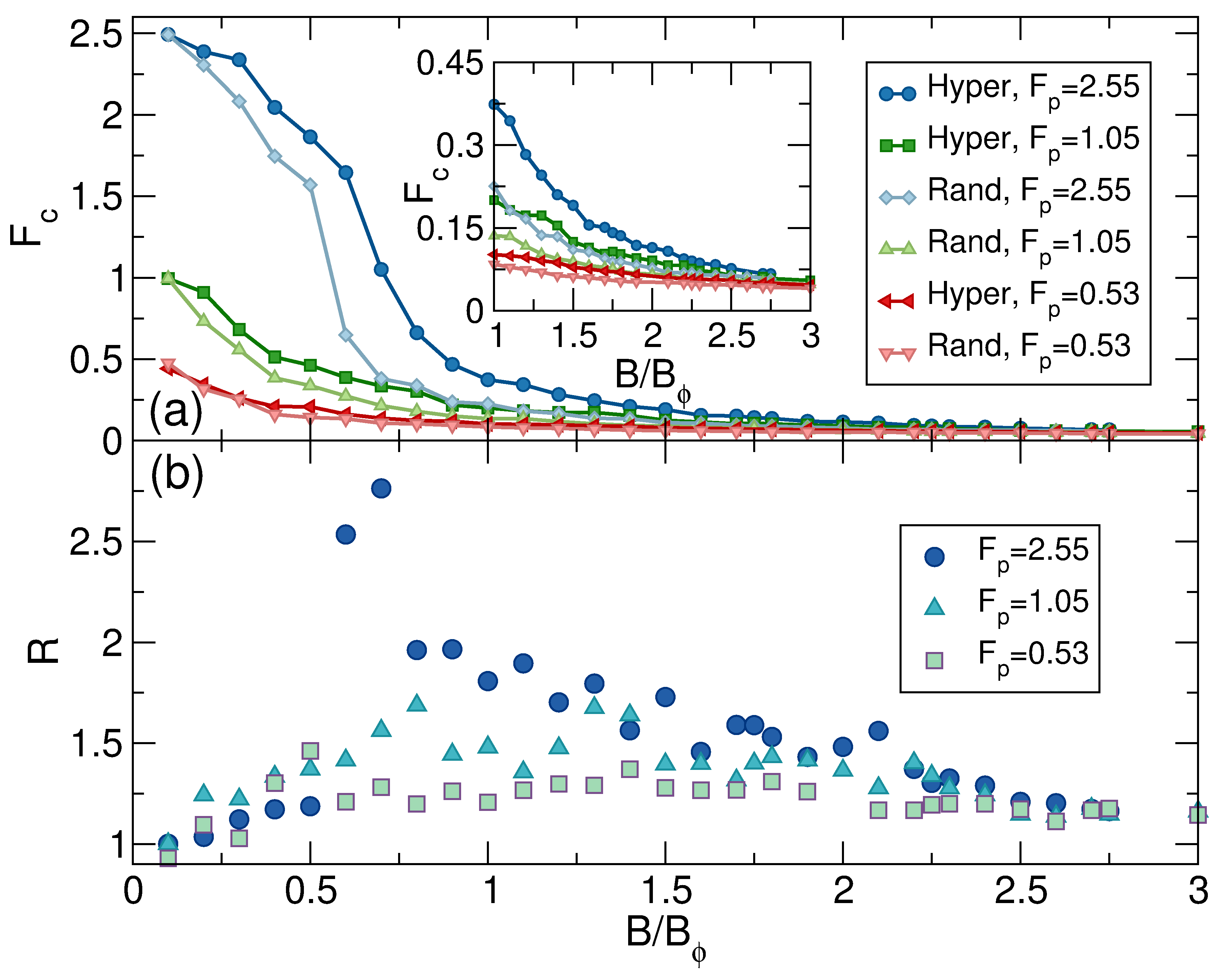}
\caption{(a) The depinning force
  $F_c$ vs $B/B_{\phi}$ for hyperuniform arrays with $F_p = 2.55$
  (dark blue circles), 1.05 (dark green squares, and 0.53 (dark red left triangles), 
  and for random pinning arrays
  with $F_{p} =2.55$, (light blue diamonds), 1.05 (light green up triangles), and $0.53$
  (orange down triangles).
  The inset shows a blow-up of the behavior at higher fields.
  (b)
  The depinning threshold ratio
  $R = F^{\rm hyper}_{c}/F^{\rm random}_{c}$ vs $B/B_{\phi}$ for
  $F_{p} = 2.55$ (dark blue circles), 1.05 (light blue triangles),
  and $0.53$ (green squares),
showing that
the pinning is consistently enhanced for the hyperuniform pinning arrays.
}
\label{fig:3}
\end{figure}

In Fig.~\ref{fig:3}(a) we plot $F^{\rm hyper}_{c}$ and $F^{\rm random}_{c}$
versus $B/B_{\phi}$
for the system in Fig.~\ref{fig:2} at varied pinning
strengths of
$F_{p} = 2.55$, 1.05, and $0.53$.
For all cases, $F_{c}$ decreases monotonically with increasing $B/B_{\phi}$, and
the hyperuniform arrays consistently have a higher $F_{c}$ than the random arrays
for a given value of $F_p$.
Figure~\ref{fig:3}(b) shows the corresponding
depinning threshold ratio $R$ versus $B/B_{\phi}$,
where $R$ approaches $R=1.0$ in the 
$B/B_{\phi}=0$ limit.
In all cases there is an enhancement of 
$F_{c}$ for the hyperuniform arrays, and
the largest enhancement occurs over the range
$0.5 < B/B_{\phi} < 2.5$.  In this regime, the $F_{p} = 2.55$ system
has some fields at which the enhancement
is as large as  $R=2.75$. At higher values
of $B/B_{\phi}$, the vortex-vortex interactions begin
to dominate over the pinning
interactions, and the differences between the hyperuniform and random arrays
are reduced.

\begin{figure}
\includegraphics[width=\columnwidth]{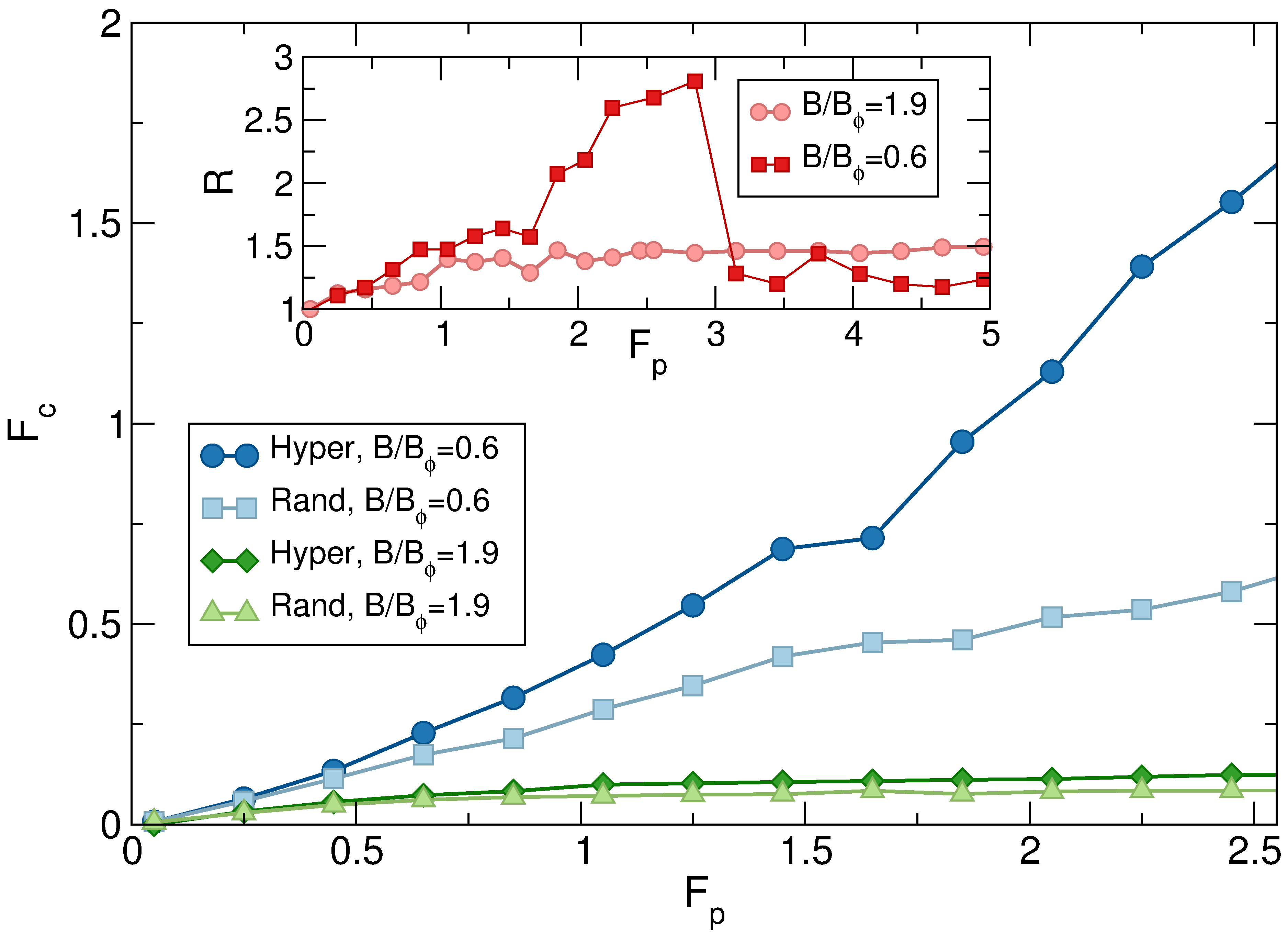}
\caption{
  The depinning force $F_{c}$ vs $F_{p}$
  for hyperuniform arrays at $B/B_{\phi}=0.6$ (dark blue circles)
  and $B/B_{\phi}=1.9$ (dark green diamonds) and for random arrays
  at $B/B_{\phi} = 0.6$ (light blue squares) and $B/B_{\phi}=1.9$ (light green
  triangles). Inset: the depinning current ratio $R$ vs $F_{p}$
  for $B/B_{\phi}=0.6$ (red squares) and $B/B_{\phi}=1.9$ (pink circles).
}
\label{fig:4}
\end{figure}

In Fig.~\ref{fig:4}
we plot $F^{\rm hyper}_{c}$ and $F^{\rm random}_{c}$ versus
pinning strength $F_{p}$ at $B/B_{\phi} = 0.6$ and $B/B_{\phi}=1.9$, while
the inset shows the corresponding $R$ vs $F_{p}$ curves.
The value of $R$ can be as
large as $R=2.75$ for $B/B_{\phi} = 0.6$,
but $R$ falls to $R=1.25$ for
higher $F_{p}$ when the pinning begins to dominate the behavior.
For $B/B_{\phi} = 1.9$,
the maximum enhancement is only $R = 1.5$,
but the enhancement remains more
robust out to higher values of $F_{p}$.

\begin{figure}
\includegraphics[width=\columnwidth]{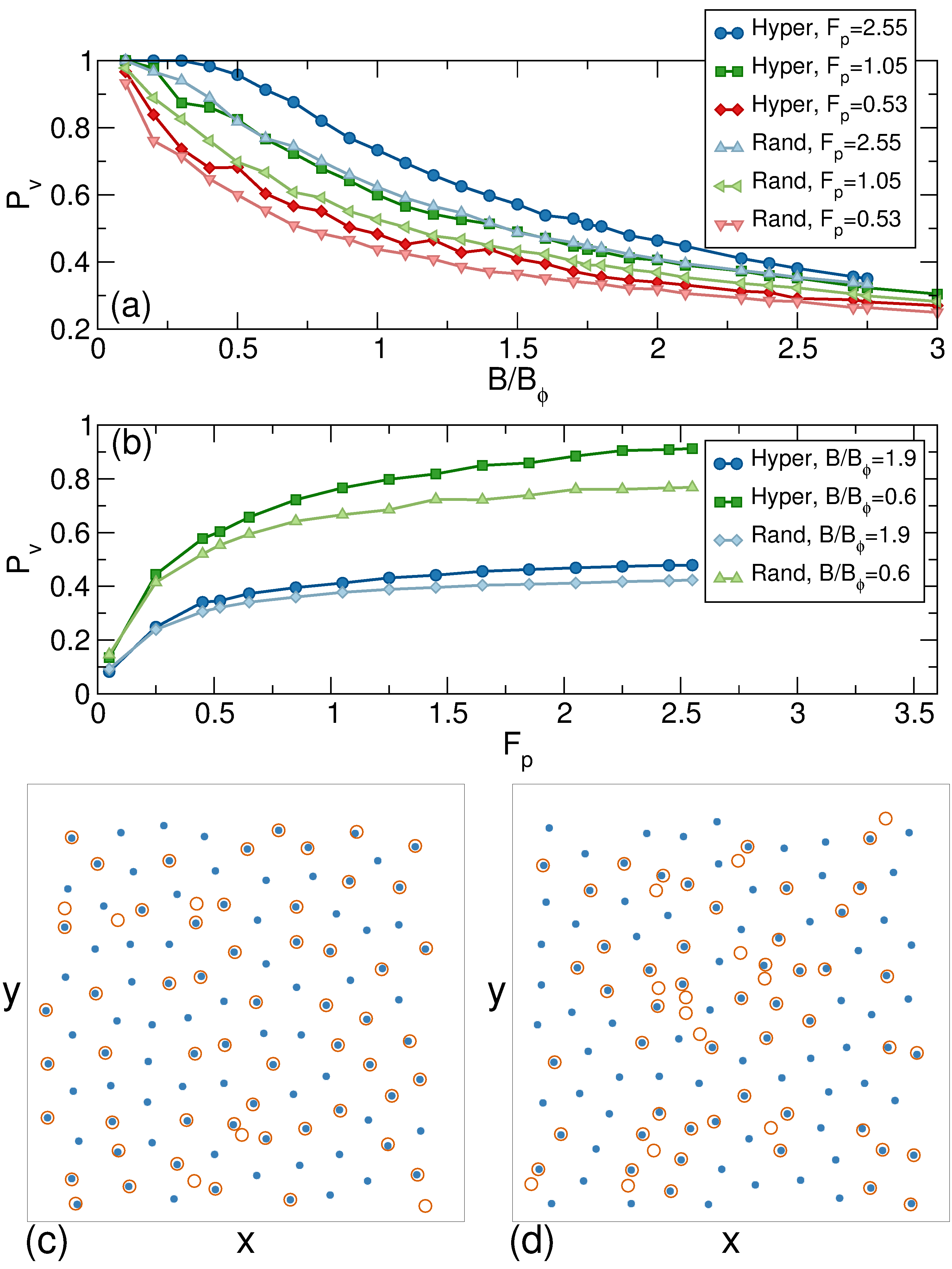}
  \caption{ (a) Fraction $P_v$ of vortices located at pinning
    sites vs $B/B_{\phi}$ for hyperuniform arrays at
    $F_{p} = 2.55$ (dark blue circles), $1.05$ (dark green squares), and $0.53$
    (red diamonds),
    and random arrays at
    $F_{p} = 2.55$ (light blue up triangles), $1.05$ (light green left triangles),
    and $0.53$ (orange down triangles),
    showing that there is a consistently higher fraction of occupied pinning
    sites in the hyperuniform arrays.
    (b) $P_v$ vs $F_{p}$ for hyperuniform arrays at
    $B/B_{\phi} = 1.9$ (dark blue circles) and $0.6$ (dark green squares) and
    random arrays at
    $B/B_{\phi} = 1.9$ (light blue diamonds) and $0.6$ (light green triangles),
    showing a similar trend.
    (c) The vortex (blue filled circles) and pinning site (orange open circles)
    locations in a small portion of the sample
    for a hyperuniform array at $F_{p} = 2.55$ and $B/B_{\phi} = 1.5$.
    (d) Vortex (blue filled circles) and pinning site
    (orange open circles) locations in a small portion of the sample for the
    random array under the same conditions showing
    that a higher fraction of pinning sites are unoccupied.
}
\label{fig:5}
\end{figure}

To better understand how the hyperuniform arrays produce enhanced pinning,
in Fig.~\ref{fig:5}(a) we plot the fraction
$P_v$ of vortices located at pinning sites versus $B/B_{\phi}$
at $F_{p} = 2.55$, 1.05, and 0.53
for the random and hyperuniform arrays,
showing that for any fixed value of $F_{p}$,
a higher fraction of vortices are located at pinning sites
for the hyperuniform array than for the random array.
In Fig.~\ref{fig:5}(b) we plot $P_v$ versus
$F_p$ for samples with $B/B_{\phi} = 1.9$ and
$0.6$, where a similar trend appears.
Figure~\ref{fig:5}(c,d) illustrates
the vortex and pinning site locations in
a small section of the sample
for $B/B_{\phi} = 1.9$ and $F_{p} = 2.55$.
Here, there are five unoccupied pinning sites in the hyperuniform array in
Fig.~\ref{fig:5}(c), while there are eleven unoccupied pinning sites in the
random array in Fig.~\ref{fig:5}(d).
In the random
array, local clumping of the pinning site positions can occur, and if a vortex is trapped
by one pinning site in such a clump, its repulsive force can screen the remaining pins and
prevent other vortices from occupying them.
The random array can also contain large spatial regions in which there are no
pinning sites,
and vortices located in these regions can flow relatively easily along
river-like channels or weak links,
depressing the value of $F_{c}$. 
In the hyperuniform array, pinning density fluctuations are suppressed,
so there is less screening of the pinning sites and a correspondingly higher
pin occupation fraction, as shown in Fig.~\ref{fig:5}.
In periodic pinning arrays, pinning density fluctuations are absent;
however, due to the symmetry of the pinning lattice,
there are easy flow directions along which vortices can form one-dimensional
easy-flow channels,
particularly at incommensurate fillings \cite{32}.
It may be possible to construct other types of hyperuniform arrays
beyond the ones we consider here which would allow for even stronger
enhancement of the pinning,
or to create a pinning lattice that is
hyperuniform along only one direction.

{\it Emergent Hyperuniformity in Vortex Systems}

\begin{figure}
\includegraphics[width=\columnwidth]{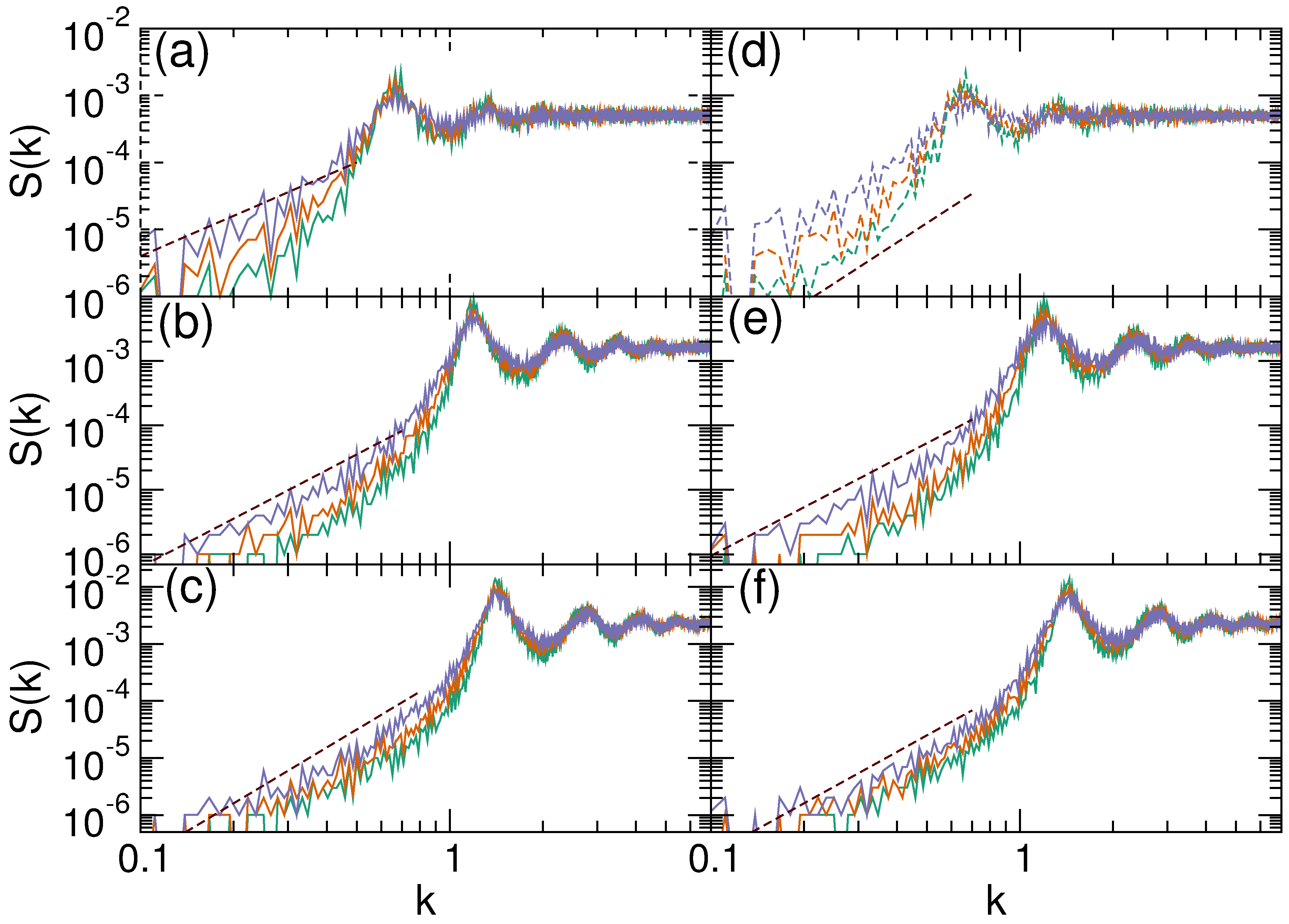}
\caption{
  (a,b,c) $S(k)$ of the vortex positions vs $k$ for $n_{p} = 0.7$ at
  $F_{p} = 0.55$ (green), $1.05$ (orange), and $2.55$ (purple)
  for a random pinning array at
  (a)  $B/B_{\phi} = 0.6$, (b) $B/B_{\phi}=1.9$, and
  (c) $B/B_{\phi}=2.7$.
  In each case $k$ goes to zero as a power law, as indicated by the
  dashed line which is a power law fit with exponent
  (a) $\alpha = 2.0$, (b) $\alpha=2.5$, and (c) $\alpha=3.25$.
  (d,e,f) $S(k)$ of the vortex positions vs $k$ for $n_{p} = 0.7$ at the same $F_p$ values
  as above for a hyperuniform pinning array at
  (d) $B/B_{\phi} = 0.6$, (e) $B/B_{\phi} = 1.9$, and (f) $B/B_{\phi} = 2.7$.
  Dashed lines indicate power law fits with exponents of
  (d) $\alpha=3.0$, (e) $\alpha=2.5$, and (f) $\alpha=3.0$.
}
\label{fig:6}
\end{figure}

\begin{figure}
\includegraphics[width=\columnwidth]{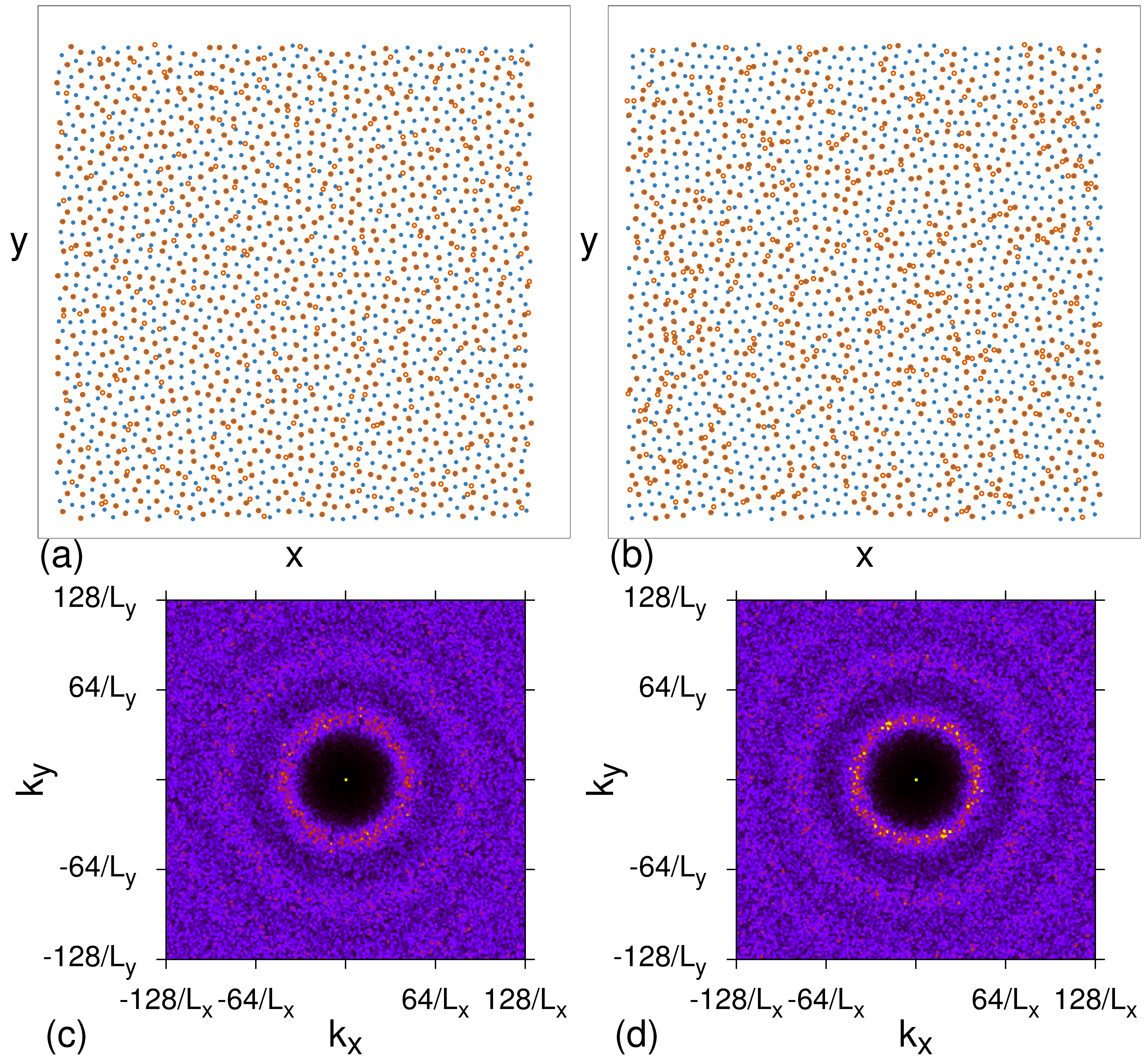}
\caption{Vortex (blue filled circles) and
  pinning site (orange open circles) location
  in the entire sample at $B/B_{\phi} = 1.9$ at $F_{p} = 2.55$ for
  (a) a hyperuniform pinning array and (b) a random pinning array.
  (c) Structure factor $S({\bf k})$ for the vortex positions in panel (a).
  (d) $S({\bf k})$ for the vortex positions in panel (b).
}
\label{fig:7}
\end{figure}

We next consider
whether amorphous vortex configurations in the
presence of random or hyperuniform pinning arrays
exhibit hyperuniformity.
As described above, 
disordered hyperuniform systems
have two identifying characteristics in the structure factor:
$S({\bf k})$ is isotropic, and it goes to zero at small $|{\bf k}|$.
In Fig.~\ref{fig:6}(a,b,c) we show $S(k)$ of the vortex configuration
for a random pinning array at $n_{p} = 0.7$ with
$F_{p} = 0.55$, $1.05$, and $2.55$
for $B/B_{\phi} = 0.6$, $1.9$, and $2.7$,
while in Fig.~\ref{fig:6}(d,e,f) we plot the same quantities for vortices
interacting with a hyperuniform pinning array.
In each case the vortices
form an amorphous structure,
as shown 
in Fig.~\ref{fig:7}(a,b) for the hyperuniform and random
arrays at $B/B_{\phi} = 1.9$ and $F_{p} = 2.55$.
The corresponding plots of $S({\bf k})$ for the {\it vortex} configurations appear
in Fig.~\ref{fig:7}(c,d), and show a ring feature in each case indicating that the vortices
are arranged isotropically.
In Fig.~\ref{fig:6}, all of the curves in all of the panels exhibit
a power law decay, with $S(k)$ going to zero as $S(k) \propto k^{\alpha}$
with $\alpha$  in the range $\alpha=2.5$ to $\alpha=3.5$, as indicated by the dashed lines.
In general, if $F_p$ is large or the vortex-vortex interactions are weak,
the vortex configurations are dominated by the configuration of the pinning sites,
so in the random pinning array system the vortices
would exhibit random characteristics, with $S({\bf k})$ going to a constant
value as $k \rightarrow 0$. 
In real superconductors, the vortex-vortex interaction strength
is non-monotonic as a function of field and temperature,
and it goes to zero as the upper critical field or critical temperature is approached.
It is therefore possible that a transition could occur
from a crystalline to a hyperuniform amorphous vortex state, followed by
a second transition to a truly 
amorphous random or liquid vortex state
as a function of increasing field or increasing temperature.
There are already numerous experimental observations of
amorphous vortex states
with and without large density fluctuations at higher
magnetic fields \cite{38,39,40,L1}, and it would be interesting to reexamine this
experimental data to determine whether the vortex configurations appear to be
random or hyperuniform.

\begin{figure}
\includegraphics[width=\columnwidth]{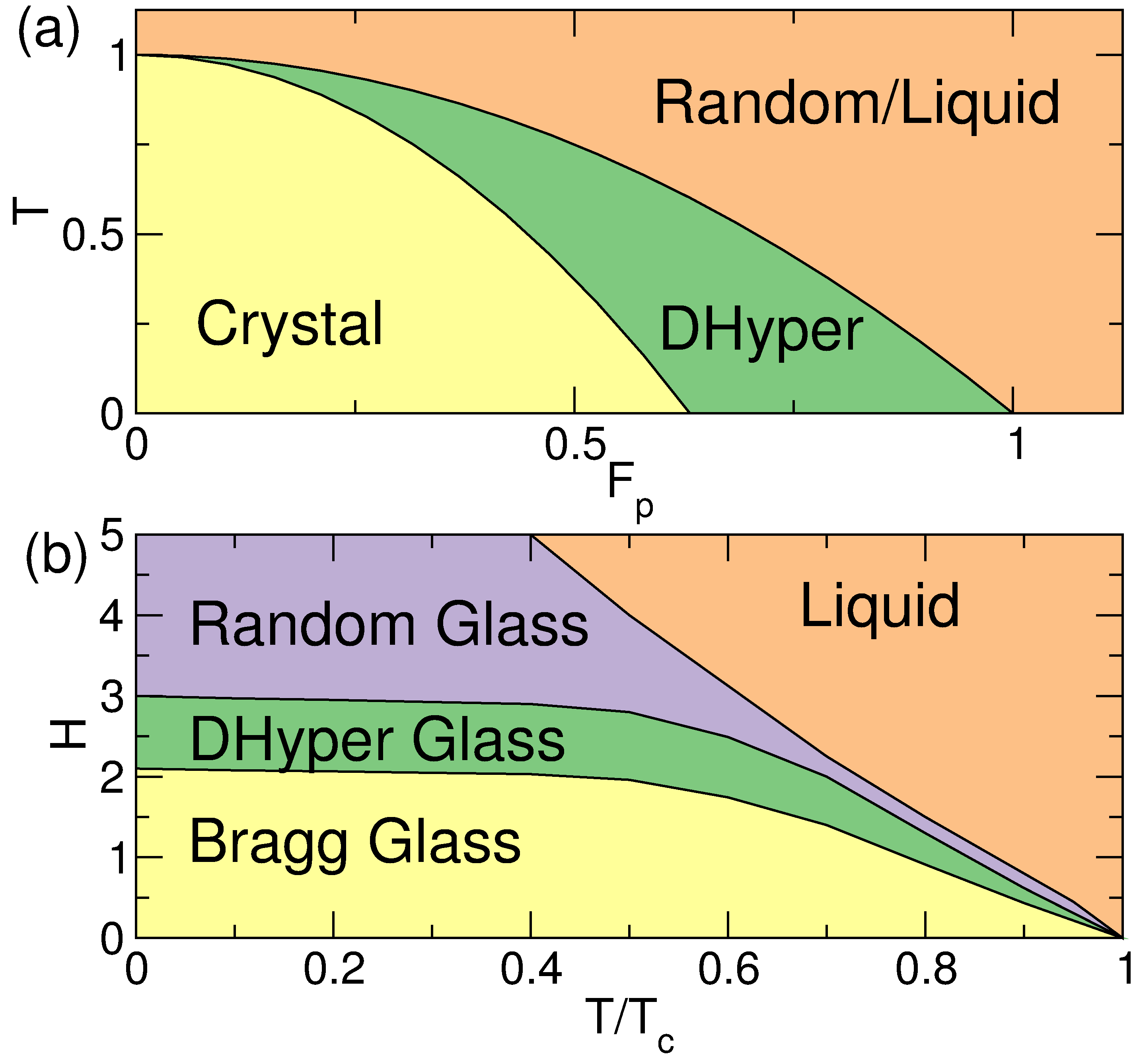}
\caption{(a) Schematic proposed phase diagram
  of temperature $T$ vs disorder strength $F_p$ for a system of repulsively  
interacting particles
in the presence of quenched disorder.
Between the crystalline and random or liquid state, there
could be a disordered hyperuniform state (DHyper).
(b) Schematic proposed modified vortex phase diagram
for a high temperature superconductor as a function of
magnetic field $H$ in arbitrary units
vs reduced temperature $T/T_c$, where $T_c$ is the critical temperature of the
material.
As a function of increasing $H$,
there is
a transition from a dislocation-free Bragg glass
into a disordered hyperuniform state, followed by transitions to
random amorphous or liquid states for high fields and temperatures.
}
\label{fig:8}
\end{figure}

In Fig.~\ref{fig:8}(a) we show a proposed
generic phase diagram for repulsively interacting particle systems as a function
of temperature versus the strength of the quenched disorder
$F_p$.  At high temperatures or for strong
disorder, the
system is amorphous
and the particle positions have random or liquid-like characteristics.
Between the crystalline state and the
random or liquid state we propose that a disordered hyperuniform state exists
for intermediate disorder strength.

In Fig.~\ref{fig:8}(b) we illustrate a proposed
variation of the vortex phase diagram for a high temperature
superconductor \cite{N1,N2,N3} in the presence of
quenched disorder.
Due to the
nonmonotonic behavior of the effective
vortex-vortex interactions as function of magnetic field and temperature,
there is a transition from a Bragg glass
state at lower fields where the vortices are dislocation-free
to an amorphous vortex glass state for increasing field or increasing temperature,
while at high fields or temperatures, a vortex liquid phase appears.
We conjecture that between the Bragg glass and the
true random amorphous vortex glass, there is a state in which
the vortex arrangement is amorphous but has hyperuniform properties.
We note that for some systems,
there can also be reentrant disordered phases at lower fields where the vortices
are far apart and the pinning becomes dominant again,
so this reentrant region could be another place to look for a
crossover from hyperuniform to random disordered vortex arrangements.
Data from imaging or neutron scattering experiments could show
whether the vortex configurations are hyperuniform at the transition between the Bragg
glass and an amorphous state, and whether at higher magnetic fields
the vortices adopt a truly random configuration.

{\it Discussion---}
We have shown that pinning sites in a disordered hyperuniform arrangement
provide
enhanced pinning compared to an equivalent number of
randomly placed pinning sites.
In hyperuniform arrays, the structure is isotropic like a liquid;
however, the density fluctuations
in the pinning site locations are
strongly reduced out to large distances, similar to
what is found in a crystal.
Random arrays are also isotropic but
can have strong density fluctuations of the type found in liquids.
In the hyperuniform pinning arrays,
we find that the probability for pinning site occupation is enhanced, while
weak links or easy flow channels are minimized due to the isotropic nature
of the pinning arrangement.
There are no symmetry directions along which easy vortex flow
can occur, unlike in crystalline pinning arrays.
We also show that amorphous vortex states in the
presence of random or hyperuniform pinning arrays
themselves exhibit hyperuniformity due
to the repulsive nature of the vortex-vortex interactions, and we propose that
there may be additional phases in the vortex phase diagram, including
both hyperuniform and random amorphous phases.
Our results should be general to the wider class of systems of
repulsively interacting
particles in the presence of either
random or hyperuniform pinning arrays, including Wigner crystals,
colloids, disordered charge systems, and skyrmions in chiral magnets.

\begin{acknowledgments}
We gratefully acknowledge the support of the U.S. Department of
Energy through the LANL/LDRD program for this work.
This work was carried out under the auspices of the 
NNSA of the 
U.S. DoE
at 
LANL
under Contract No.
DE-AC52-06NA25396 and through the LANL/LDRD program.
\end{acknowledgments}

\end{document}